# A Robust Model-free Approach for Rare Variants Association Studies Incorporating Gene-Gene and Gene-Environmental Interactions

Ruixue Fan and Shaw-Hwa Lo[*]

Department of Statistics, Columbia University, New York, NY 10027, USA
*Corresponding author.

Email addresses:
RF: rf2283@columbia.edu
SHL: slo@stat.columbia.edu

## Abstract

Recently more and more evidence suggests that rare variants with much lower minor allele frequencies play significant roles in disease etiology. Advances in next-generation sequencing technologies will lead to many more rare variants association studies. Several statistical methods have been proposed to assess the effect of rare variants by aggregating information from multiple loci across a genetic region and testing the association between the phenotype and aggregated genotype. One limitation of existing methods is that they only look into the marginal effects of rare variants but do not systematically take into account effects due to interactions among rare variants and between rare variants and environmental factors. In this article, we propose the *s*ummation of *p*artition *a*pproach (SPA), a robust model-free method that is designed specifically for detecting both marginal effects and effects due to gene-gene (G×G) and gene-environmental (G×E) interactions for rare variants association studies. SPA has three advantages. First, it accounts for the interaction information and gains considerable power in the presence of unknown and complicated G×G or G×E interactions. Secondly, it does not sacrifice the marginal detection power; in the situation when rare variants only have marginal effects it is comparable with the most competitive method in current literature. Thirdly, it is easy to extend and can incorporate more complex interactions; other practitioners and scientists can tailor the procedure to fit their own study friendly. Our simulation studies show that SPA is considerably more powerful than many existing methods in the presence of G×G and G×E interactions.

## Introduction

Despite of the success of large scale biological studies such as GWAS in discovering many disease variants, most of which are common variants with minor allele frequency (MAF) greater than 0.05, for diabetes, heart disease, Alzheimer disease, etc., the variants identified thus far confer relatively small risk, explain a small fraction of familial clustering, and add little practical value in disease prediction. The issue of so-called "missing heritability" has been a serious concern that has attracted considerable attention and discussion recently. [1,2,3,4,5] A number of explanations have been suggested for this phenomenon including: (1) an as-yet undiscovered larger set of variants of smaller effects, (2) rare variants with larger effects that may be eluding the current GWAS, (3) unaccounted effects, due to gene-gene (G×G) and gene-environment (G×E) interactions, (4) undetected structure effects including copy number variations (CNVs), and (5) over-estimated heritability.[6,7,8,9,10,11] This article presents a simple yet easy-to-extend method to address issues (2) and (3).

In genetic association studies, the 'common-disease common-variants' (CDCV) model states that common diseases are caused by common variants with MAFs greater than 5% or 1%. [12] However, recently more and more evidence supports the alternative 'common-disease rare-variants' (CDRV) hypothesis which claims that complex disorders are caused by multiple rare variants with MAF < 1%. [13,14] Unlike common variants that do not affect protein function directly, most rare variants are missense mutations in promoter region or protein coding regions and they are capable of altering gene expression level, changing amino acids sequence and affecting protein-protein interactions. [15,16] Furthermore, rare variants may have higher odds ratios (above 2), compared with small odds ratios (1.1~1.5) of common variants. [17] Therefore, the investigation of rare variants will help researchers further understand the disease etiology and may provide new insights into medical treatments. With the development and commercialization of next generation sequencing technologies, large number of SNPs with low frequencies can be detected in a relatively short time and at relatively low cost. [5] In the near future, whole-genome sequencing will become possible for large numbers of individuals, and, as a result, large amounts of sequence data with rare variants will be generated. Methods that are capable of detecting these casual variants are very much in need.

Due to the low frequencies and large number of rare variants, traditional single-marker association tests that have worked well for common variants will in general lack power for rare variants. [18] In recent years, several statistical methods have been developed based on collapsing rare variants in a specific region of interest, e.g. a gene or genes from a specific pathway, followed by performing a region-based test rather than individual tests for each variants. The Combined Multivariate and Collapsing (CMC) method proposed by Li and Leal [19] tests whether the proportions of rare variants carriers in cases and controls are significantly different. The weighted sum (WS) method by Madsen and Browning [20] is designed to weight variants according to their estimated frequencies in controls, so that less frequent variants receive higher weights compared with more common variants. Instead of using the conventional cutoff values 0.05 or 0.01 to define rare variants, Price et al. [21] proposed to choose a variable threshold (VT) that gives an optimal testing power. Ionita-Laza et al. [22] developed a replication-based (RB) approach, also based on a weighted-sum statistic, that can be more powerful in the presence of both risk and protective variants. Wu et al. proposed a sequence kernel association test (SKAT) that is a score-based variance component test. [23] SKAT uses a linear weighted kernel $K(G_i, G_{i'}) = \sum_{j=1}^{K} w_j G_{ij} G_{i'j}$ to measure the similarity between individuals $i$ and $i'$ ($K$ is the number of markers and $w_j$ is the weight of SNP $j$). A weighted quadratic kernel $K(G_i, G_{i'}) = \left(1 + \sum_{j=1}^{K} w_j G_{ij} G_{i'j}\right)^2$ was also proposed in [23] to account for both main effects and genetic interaction effects but it was not systematically studied. Many alternative methods that have also been proposed can be considered variations of these approaches. [24,25,26]

**Why another approach?** The aforesaid methods have been shown to work well in different simulated models (mostly with marginal effects only). However, all these tests only consider marginal effects from rare variants and they do not systemically address the issue of interactions among rare variants (G×G), or between rare variants and covariates, such as environmental factors (G×E). Therefore, additional statistical methods are needed to generate scientific knowledge on the etiology of complex diseases where interactions among genetic, biological and environmental variables work together to produce a phenotype. In this article, we propose the *summation of partition approach* (SPA), a robust model-free method that not only tests the marginal effects of rare SNPs but also naturally incorporates G×G and G×E interactions. As with existing methods, SPA is based on aggregating information across rare variants in a region of interest. We shall demonstrate the power of SPA and compare with existing methods for both dichotomous and quantitative phenotypes. Simulation studies show that in disease models without interactions, the performance of SPA is comparable to or even better than the most competitive existing method in current literature, and in the presence of G×G interactions, SPA substantially outperforms all the other methods. Another advantage of our procedure is its simplicity and extensibility. We also demonstrate in this article how to incorporate an environmental factor in the proposed framework and show that the augmented test score is powerful in detecting G×E interactions. Similar approaches can be taken to account for interactions with common variants or other covariates. In addition, we compare the proposed method with several existing tests on the dataset provide by Genetic Analysis Workshop 17 (GAW17) and find that SPA is robust for detecting different genes. When large volumes of datasets with rare variants become available in the near future, the proposed procedure will become a powerful tool to detect complicated interaction effects in various genetic regions and it will help us to better understand the mechanisms of complex human diseases.

## Materials and Methods

To better understand the motivation and rational behind SPA, we briefly review a general framework that has been adopted for detecting common variants with interactions. A core element in this framework is the influence score *I* derived from what we now know as the Partition Retention (PR) method. [27] Several forms and variations were associated with the PR method before it was finally coined this name in 2009.

**A General Framework Used for Detecting Common Variants**
We demonstrate a basic tool adopted by our method. Suppose there are *n* subjects with a response variable *Y* and *K* discrete explanatory variables $\{X_1,…, X_K\}$. If each $X_i$ can take three discrete values, we generate a partition $\Pi$ with $3^K$ non-overlapping partition elements. Let $n_i$ be the number of subjects in partition *i*, $\overline{Y}_i$ the average response for subjects in partition *i*, and $\overline{Y}$ the average response from all subjects. An influence measure between the response and the predictors is defined as:

$$I(X_1,…,X_K) = \sum_{i \in \Pi} n_i^2 \left(\overline{Y}_i - \overline{Y}\right)^2$$

It has been shown that under the null hypothesis that none of the predictors has influence on *Y*, the normalized *I*, $I/(n\sigma^2)$ ($\sigma^2$ denotes the variance of *Y*) is asymptotically distributed as a weighted sum of $\chi^2$ random variables of 1 degree of freedom each such that the total weight is less than 1. [27] The main structure of this measure is the partition formed by the $K$ discrete variables with $3^K$ partition elements each containing non-overlapping observations. This influence measure captures any discrepancy between the conditional mean and the grand mean of $Y$ and thus is able to detect $X$-$Y$ association regardless of the structure of dependence. It can be easily generalized to any discrete random variables with finite number of outcomes.

In case-control studies, the influence measure can be rewritten as:

$$I = \sum_{i \in \Pi} n_i^2 \left( \hat{p}_i^D - \frac{N_A}{N_A + N_U} \right)^2$$

where $N_A$ is the number of affected individuals, $N_U$ is the number of unaffected individuals, and $\hat{p}_i^D$ is the proportion of cases in partition $i$. Several variations of this partition-based method have been successful at identifying influential common variants and their interactions in human diseases, such as Rheumatoid Arthritis [28,29,30] and breast cancer [31,32]. Its success in detecting common variants relies on the essence that many partition cells contain more than singleton subjects, however, this property will diminish for rare variants due to their extremely low frequencies. To effectively deal with rare variants, we need to modify the partition procedure properly to accommodate for the sparseness, which can be achieved by the proposed *s*ummation of *p*artition *a*pproach (SPA). We introduce below several test statistics of SPA, including the marginal test score $I_1$, G×G interaction score $I_2$, and G×E interaction scores $I_2^*$.

**Rare Variants Marginal Association Score $I_1$**

The general framework mentioned above can be extended to rare variants association analysis for both dichotomous and continuous phenotypes.

In population-based case-control studies, suppose there are $N$ unrelated individuals, among which $N_A$ are cases and $N_U = N - N_A$ are controls. The region of interest $G$ contains $K$ rare variants and the genotype of the $j^{th}$ individual is denoted $(X_1^{(j)},...,X_K^{(j)})$. Each $X_i^{(j)} (i = 1,...,K)$ can take values 0, 1 or 2, indicating the number of rare variants at this position. The SPA test score $I_1$ that accounts for all marginal information contributed by these $K$ rare SNPs is defined as

$$I_1 = \sum_{i=1}^{K} n_i^2 \left( \hat{p}_i^D - \frac{N_A}{N_A + N_U} \right)^2$$

where $\hat{p}_i^D$, for the $i^{th}$ SNP, is the fraction of all observed rare variants that are from cases, and $n_i$ is the total number of $i^{th}$ rare variant observed in all subjects.

For continuous traits, $I_1$ is defined as

$$I_1 = \sum_{i=1}^{K} n_i^2 \left( \overline{Y}_i - \overline{Y} \right)^2$$

where $\overline{Y}_i$, for the $i^{th}$ SNP, is the averaged response for subjects bearing at least one rare variant, $\overline{Y}$ is the averaged response from all subjects and $n_i$ is defined as above. Different from the original influence measure, $I_1$ recognizes the partition elements formed by individual SNP and hence the partitions from different SNPs are not non-overlapping any more; therefore, $I_1$ does not suffer from the sparseness of rare variants. Under the null hypothesis of no influence, the differences between $\hat{p}_i^D$ and $\frac{N_A}{N_A + N_U}$ for dichotomous traits (or between $\overline{Y_i}$ and $\overline{Y}$ for continuous traits) for all $i$ are small, so a large $I_1$ value indicates that some rare variants in the region might be associated with the disease phenotype. Additionally, since each term of $I_1$ is the squared difference between the conditional average and the grand average, it can detect both directions of departure from the expected difference zero, which renders $I_1$ the ability to capture the association even in a region with both risk and protective rare variants. Unlike PR's influence measure $I$, the statistical property of $I_1$ is more complicated to obtain since the dependence between partition cells created by different SNPs will not asymptotically disappear even under the null hypothesis of no influence. Therefore, in our analyses we will rely on the method of permutation to assess its statistical significance.

**Rare Variants G×G Interaction Association Score $I_2$**
In order to increase the power of detecting the genotype-phenotype associations as well as to elucidate the biological pathways that underpin disease, more and more attentions have been given to the identification of interactions between SNP loci. [33,34,35] A limitation of $I_1$ is that it considers little interactions among rare SNPs. From the general framework, we propose a second SPA test score $I_2$ that evaluates G×G interactions among rare variants.

As the genotype at each SNP position can take 3 values, in theory we are facing a maximum of $3^K$ partition elements for all levels of interactions. However, due to the low frequencies of rare variants, the higher order (>2) interaction information among rare SNPs in current sample size will be small. For example, if the sample size is 1,000 and the SNP frequency is 0.01, the expected number of observing one specific rare variants triplet is $1{,}000 \times 0.01^3 = 10^{-3}$. If a region contains 20 independent rare SNPs, the expected total number of rare variants triplets would be $\binom{20}{3} \times 0.001 = 1.14$, which provides very low signal for 3-way interaction detection. Therefore, for current sample size, we only consider an influence measure that takes into account 2-way interactions among rare variants. For a pair of rare SNPs $i$ and $j$, we consider three aggregated cells: individuals with rare variants only on SNP $i$ (denoted $mM$), individuals with rare variants only on SNP $j$ (denoted $Mm$) and individuals with rare variants on both SNPs (denoted $mm$). Note that we do not consider the cell $MM$ where individuals have no rare variant at either position. For dichotomous trait, the SPA test score $I_2$ for G×G interaction is defined as

$$I_2 = \sum_{i \geq 1, j > i}^{K} n_{ij}^2 \left[ \left( \hat{p}_{ij,mM}^D - \frac{N_A}{N_A + N_U} \right)^2 + \left( \hat{p}_{ij,Mm}^D - \frac{N_A}{N_A + N_U} \right)^2 + \left( \hat{p}_{ij,mm}^D - \frac{N_A}{N_A + N_U} \right)^2 \right]$$

where $n_{ij}$ is the number of subjects who have at least one rare variant in either SNP ($i$ or $j$), $\hat{p}^D_{ij,mM}$ is the fraction of subjects that are cases in partition $mM$, $\hat{p}^D_{ij,Mm}$ is that fraction in partition $Mm$, and $\hat{p}^D_{ij,mm}$ in partition $mm$. For quantitative trait, $I_2$ is defined as

$$I_2 = \sum_{i \geq 1, j > i}^{K} n_{ij}^2 \left[ \left( \bar{Y}_{ij,mM} - \bar{Y} \right)^2 + \left( \bar{Y}_{ij,Mm} - \bar{Y} \right)^2 + \left( \bar{Y}_{ij,mm} - \bar{Y} \right)^2 \right]$$

where $\bar{Y}_{ij,mM}$ is the average response for individuals in partition $mM$, $\bar{Y}_{ij,Mm}$ in partition $Mm$, and $\bar{Y}_{ij,mm}$ in partition $mm$. If two rare variants have interactions, the difference between the conditional average and the unconditional average will be large, leading to a large $I_2$ value. Again, permutation is used to evaluate the significance of the test statistic $I_2$. Even though $I_2$ only considers 2-way interaction, it can be easily extended to include higher-order ($\geq 3$) interactions by generating partitions based on $m$-tuples ($m \geq 3$) of rare SNPs.

**Adaptive Test Score $p^*$**
When we are unclear whether G×G interaction is involved in the onset of disease, we propose an adaptive score $p^*$ that is a compromise between $I_1$ and $I_2$. We first evaluate the significance of $I_1$ and $I_2$. Then the adaptive test score is defined as:
$$p^* = \min(p(I_1), p(I_2))$$
where $p(I_1)$ and $p(I_2)$ are the p-values of $I_1$ and $I_2$ separately. We evaluate the significance of $p^*$ by permutation.

**Rare Variants G×E Interaction Association Score $I_2^*$**
Increasing evidence has shown that gene and environmental (G×E) interactions are widely involved in the etiology of complex diseases, including diabetes, cancer and psychiatric disorders [36,37,38,39,40]. Conventional methods to detect G×E interactions are mostly based on regression models, which will lose power for rare variants. SPA can be easily extended to incorporate covariates, such as environmental factors in the testing procedure, considering both the environmental marginal effect and the G×E interaction information. Here we focus on case-control study design. Suppose an environmental factor $E$ has $J$ levels. The SPA test score for detecting the effect of the environmental factor is expressed as:

$$I_2^* = \sum_{j=1}^{J} \sum_{i=1}^{K} n_{i,j}^2 \left( \hat{p}^D_{i,j} - \frac{N_A}{N_A + N_U} \right)^2$$

where $\hat{p}^D_{i,j}$ is the fraction of rare variants at position $i$ on level $j$ that are from cases, and $n_{i,j}$ is the total number of $i^{th}$ rare variants observed at level $j$. $I_2^*$ is a modification of $I_1$ by building additional overlapping rare variants partition cells to $J$ non-overlapping partitions created by the environmental factor. The significance of $I_2^*$ is evaluated by permutation. We propose two permutation strategies: (1) global permutation that permutes the phenotype among all individuals; and (2) local permutation that permutes the phenotype within each stratum of the environmental factor. Both permutation strategies are investigated in our study.

**Simulation Scheme**
We simulated several scenarios for the purpose of evaluation and comparison of our test scores with several existing rare variants association methods. The genotype consists of 20 independent rare variants in each scenario. Scenario 'Null-1' is a 'null model' where none of the 20 variants affects the phenotype. For dichotomous traits, the phenotypes are determined by the baseline penetrance only. This is the null setting for $I_1$, $I_2$, $p^*$ and $I_2^*$ with global permutation. In scenario 'Null-2', the dichotomous outcomes are affected by the environmental factor. 'Null-2' is the null setting for $I_2^*$ with local permutation.

For empirical power comparisons, we generate three different sets of simulations. The first set of simulations are marginal effect models, in which the MAF of all SNPs are uniformly distributed between 0.0001 and 0.01. In scenario 1, 5 out of the 20 rare SNPs are risk SNPs and the effect size is constant. Scenario 2 is similar to scenario 1 except that the risk effect is negatively correlated with MAF. Scenario 3 has 5 protective variants and 5 deleterious variants with effect size negatively correlated with MAF. The second set of simulations contains 2-way G×G interaction between rare variants, with MAF 0.01 for all 20 SNPs. In scenario 4, 50% of the SNPs (10 out of 20 SNPs) have interaction effects. Scenario 5 is similar to scenario 4 but 75% of the SNPs are involved in G×G interactions. Both main effect and G×G interaction effect exist in scenario 6. The third set of simulation models involves G×E interaction effects with a binary environmental factor. Scenario 7 has positive G×E interaction effects and environmental marginal effect; scenario 8 has both positive and negative G×E interaction effects. Logistic regressions or linear regression was used to generate dichotomous or quantitative phenotypes. 1,000 repetitions were simulated for each scenario with four different sample sizes, each having equal number of cases and controls. Detailed simulation models are provided in Table S1 in file S1.

# Results

We compared the power of SPA test scores $I_1$, $I_2$ and $p^*$ with existing methods: CMC, WS, VT, RB SKAT (with the weighted linear kernel) and SKATint (a modified SKAT score with the weighted quadratic kernel) in a series of simulation scenarios, including marginal effect models and G×G interaction effect models for both dichotomous traits and continuous traits. RB only deals with binary outcomes, so it is not included in our analysis for continuous traits. We also evaluated the power of $I_2^*$ in G×E interaction effect models for dichotomous traits. (See *Material and Methods* for details of simulation models; numerical results from our simulation studies are presented in Table S2-S7 in file S1.)

**Type I Error of $I_1$, $I_2$ and $p^*$**
The empirical type I error rates for $I_1$, $I_2$ and $p^*$ are presented in Table 1 for nominal levels α=0.05 and α=0.01 with four different sample sizes: 600, 1000, 1500 and 2000. The results show that $I_1$, $I_2$ and $p^*$ are well controlled at both significance levels for either dichotomous or continuous trait, even when the sample size is small, indicating that the proposed tests are valid methods. Additional results of type I error for competing methods are presented in Fig. S1 in file S1.

**Power Comparison in Marginal Effect Models for both Dichotomous and Continuous Traits**

We compare the power of $I_1$, $I_2$ and $p^*$ with competing methods in three marginal effect models when (1) only risk variants exist and the effect size is constant, (2) only risk variants exist and the effect size is negatively correlated with MAF, or (3) a mixture of risk and protective rare variants exists.

In all three marginal effect scenarios, the performance of $I_1$ and SKAT are comparable and they are both superior to the other tests (Fig. 1, 2 and 3). For dichotomous traits, $I_1$ is the most powerful method, followed by SKAT and $p^*$. For continuous traits, SKAT and $I_1$ are most competitive; both of them are more powerful than the other methods. The power of the adaptive score $p^*$ is very close to $I_1$; $p^*$ is much more powerful than CMC, WS, VT and RB. In addition, $I_1$ and $p^*$ are quite robust to different simulation scenarios, even in the presence of a mixture of risk and protective variants, while CMC, WS and VT suffer substantial power loss when causal rare variants have opposite effects (Fig. 3). It is worth noting that although $I_1$ does not intentionally highlight less frequent variants by giving them higher weights, it is still the most powerful (for dichotomous trait) or the second most powerful (for quantitative traits) method even in scenarios where the effect size is negatively correlated with MAF, showing that its good performance is intrinsic and is not driven by a specific weighting scheme. The test score $I_2$ does not show a high power in these marginal effect models as it is designed to detect G×G interaction effects but not the marginal effect.

**Power Comparison for G×G Interaction Effect Models for both Dichotomous and Continuous Traits**

We evaluated the power of different methods in two G×G interaction effect models (scenarios 4 and 5). The advantage of the G×G interaction association score $I_2$ over all the other methods is apparent for both dichotomous and continuous traits (Fig. 4 and Fig. 5). For dichotomous traits, when the sample size is large, the power of $I_2$ is substantially higher than all the other methods. For continuous traits, $I_2$ is uniformly the most powerful method for all sample sizes; for example, when the sample size is 2000, $I_2$ is 38% more powerful than SKATint at α = 0.01. Moreover, the adaptive score $p^*$ has a power that is just slightly less than $I_2$, and $p^*$ is substantially more powerful than the rest. On the other hand, VT, WS and CMC suffer from significant loss of power in the presence of complicated G×G interaction effects.

We also examine the scenario in which the phenotypes are influenced by both genetic marginal and G×G interaction effects (scenario 6). Here the marginal effect is set to be small so that it will not mask the interaction effect. $I_2$ is still consistently the most powerful test and $p^*$ is the second best, followed by SKATint (Fig. 6). For continuous traits with sample size 2000, $I_2$ is 29% more powerful than SKATint, and $p^*$ is 28% more powerful than SKATint at α = 0.01.

**Type I Error and Power of $I_2^*$ for Dichotomous Trait**

For the G×E interaction score $I_2^*$, we investigated its type I error and power for dichotomous trait using two permutation strategies – global permutation and local permutation (see *Materials and Methods*), denoted by $I_2^*$-*Global* and $I_2^*$-*Local* respectively. As $I_2^*$ considers both the genetic and environmental marginal effects as well as G×E interaction effect, $I_2^*$-*Global* is appropriate for testing the null hypothesis of no association at all (no G marginal, E marginal or G×E interaction effects), and $I_2^*$-*Local* is appropriate for testing the null hypothesis of no E marginal effect.

The type I error of $I_2^*$ are evaluated for two null hypotheses. The first null hypothesis (null-1) assumes the dichotomous traits are completely determined by the baseline penetrance. The second null hypothesis (null-2) assumes that the phenotypes are affected by environmental marginal (E marginal) effect. Table 2 presents the type I error of $I_2^*$-*Global* and $I_2^*$-*Local* in these two null settings. In null-1, both $I_2^*$-*Global* and $I_2^*$-*Local* are correctly controlled at levels α=0.05 and 0.01. In null-2, $I_2^*$-*Local* still hits the target level while $I_2^*$-*Global* has significant higher values. This is because $I_2^*$-*Global* is able to test any effect from genetic or environmental factors, including the E marginal effect; hence the results of $I_2^*$-*Global* in null-2 are indeed the *power* of $I_2^*$-*Global* in the presence of E marginal effect. On the other hand, $I_2^*$-*Local* removes the E marginal effect, so it shows the correct type-I error in both null-1 and null-2.

Two scenarios are considered to compare the power of $I_2^*$-*Global*, $I_2^*$-*Local* and competing methods when (1) the phenotypes are affected by E marginal effect and positive G×E effect, (2) the phenotypes are affected by E marginal effect and both positive and negative G×E effects. In computation, SKAT and SKATint regress the phenotype on the environmental factor when calculating the test statistic [23]. $I_2^*$-*Global* and $I_2^*$-*Local* use the environmental factor as in their definition. All the other methods work on the phenotype and the genotype directly. The results show that $I_2^*$-*Global* has much higher power than all the other tests because it takes into account both E marginal and G×E interaction effects, and $I_2^*$-*Local* outperforms all the remaining methods that do not consider G×E interaction effects (Fig. 7).

**Application to the GAW17 Dataset**

The genetic analysis workshop 17 (GAW17) provided genotypes of 3,205 autosomal genes on 697 individuals from the 1000 Genome Project. A dichotomous phenotype was simulated from a linear model using SNPs from 34 genes and most causal SNPs were rare variants. A total of 200 simulation replicates were carried out and the genotype was held fixed for all replicates. See [41] for more details of the simulation model. Here we chose to re-analyze two causal genes *FLT1* and *ANRT*. In the workshop, *FLT1* has been shown to exhibit a strong signal in many well-known methods while *ARNT* could not be identified by any existing approach. For both genes, an upper frequency of 0.05 was used as the MAF cutoff to define rare variants and only nonsynonymous SNPs were examined. We computed the power of our test scores and competing methods using all 200 replicates. Power was calculated as the proportion of replicates with p-value less than 0.05 out of the 200 simulations. As shown in Table 3, $I_1$ was fairly robust for detecting both genes. For *FLT1*, two count-based collapsing methods – CMC and WS are most powerful, followed by VT and $I_1$. For *ARNT*, $I_1$ is substantially more powerful than the other methods – its power is 47% higher than the second best method SKAT. Given that the simulated model is a simple additive linear model with genetic marginal effects only, methods considering G×G interactions, including $I_2$ and SKATint, do not have apparent advantages in power gain for detecting either *FLT1* or *ARNT*.

**Computation Time**
The computation time of $I_1$, $I_2$ and $p^*$ depends on the sample size, the number of variants and the number of permutations. On a 2.66GHz laptop with 4GB memory, to reach a significance level of $10^{-4}$, the computation times to analyze a region with 20 SNPs for 600, 1000, 1500 and 2000 individuals are 3, 5, 7, 10 sec for $I_1$, and are about 1000, 1400, 1900, 2500 sec for $I_2$.

# Discussion

We propose here the summation of partition approach (SPA), a flexible robust model-free framework for rare variants association analysis that incorporates both G×G and G×E interactions. The proposed SPA test scores create partitions from individual SNP and combine the information across all rare variants in a region of interest. $I_1$ is designed to detect marginal effects of rare variants and $I_2$ is designed to capture the G×G interaction effects among rare variants. In various marginal effect models, $I_1$ is more powerful than most approaches examined in our study. Its performance is comparable to SKAT, which is regarded as the most competitive existing method. In G×G interaction models or in the scenario with both marginal and G×G interaction effects, $I_2$ is superior to all the other methods in terms of detection power. The adaptive score $p^*$ is a compromise between $I_1$ and $I_2$ and has the advantage of both test scores. Its performance is just a little shy of the better of the two scores $I_1$ and $I_2$, for both marginal effect models and interaction effect models. Therefore, $p^*$ is a self-tuning adaptive score that is able to gain power automatically regardless of the simulation scenario. In practice when we have no clue of how the genotype affects the phenotype, we suggest to use the adaptive score $p^*$. A significant p-value of $p^*$ indicates a potential true signal from either marginal or interaction effects of rare variants. In our study, we focus on the situation with 20 rare SNPs. If the SNP number changes to 30, the simulation results (Fig. S2 in file S1) are qualitatively similar in that $I_1$ is the most powerful in marginal effect models and $I_2$ is the most powerful in interaction effect models.

$I_2^*$ is an augmented score of $I_1$ that incorporates covariates. It can be used to test the hypothesis of no association at all (neither G marginal, E marginal nor G×E effect) using 'global permutation' or to test the hypothesis of no E marginal effect using 'local permutation'. By 'local permutation', $I_2^*$ removes the marginal effect of the environmental factor while still captures variations of the genetic effect at different levels of the environmental factor. In a similar fashion, covariates can be incorporated into $I_2$ and the resulting augmented score could be used to detect E×G×G 3-way interactions between an environmental factor and two rare variants.

$I_2^*$ can also be used to test the interaction effect between common and rare variants if one treats the common variant as an environmental factor. It can be further extended to detect 3-way interactions among the environmental factor, common and rare variants by building additional overlapping partitions based on rare variants on top of the non-overlapping partition cells generated by the environmental factor and the common variant. A global permutation can detect both main and interaction effects of these factors, and a local permutation that permutes the phenotype within each non-overlapping partition cell will capture the E×common×rare 3-way interaction effect.

$I_2^*$ deals with categorical covariates naturally. In order to handle continuous covariates, such as age, height and BMI, we suggest taking the discretization approach that divides continuous variables into distinct buckets. These 'pseudo-categorical' variables generated by discretization can be applied to $I_2^*$ directly. In practice, we usually set the number of buckets to be 2~5 and the results are quite satisfactory. Moreover, a new influence measure dealing with continuous covariates directly is under preparation.

The insight of SPA is similar to the partition retention (PR) influence measure as in [27]. The PR method generates non-overlapping partition elements over the sample space and assigns each partition cell a weight that is proportional to the probability of falling into that cell. Its success in detecting influential variables relies on the essence that weights are not too small for all partition elements, especially for those cells that generate signals. Therefore, the PR method may lose power for rare variants association studies as the partition cells with true signals will have very low weights due to the extremely low frequencies of rare variants. SPA differs from the PR method by creating overlapping partition elements to avoid the sparseness and to boost the signal from rare variants.

The information measure $I_1$ can be viewed as a special case of

$$I_1 = \sum_{i=1}^{K} w_i n_i \left( \hat{p}_i^D - \frac{N_A}{N_A + N_U} \right)^2$$

where $\{w_i\}$ are weights that sum to 1. Weights can be defined in various ways. The inherent choice we take here is $w_i = n_i \Big/ \sum_{i=1}^{K} n_i$. If external information is available on possible effects of a rare variant to disease, it is straightforward to incorporate such information in our test approach by tuning the weight. Some commonly used weights are based on (1) MAF of the variant as in [20]; or (2) externally-defined weights such as predictions from SIFT and PolyPhen, as suggested by Price et al. [21]. In our study, even though we do not incorporate the weight information, SPA is still superior over the other methods. We believe that after tuning the weight, SPA will exhibit a better performance.

Population stratification has been shown to be an important problem for common variant association analysis. For rare variants, this problem is more likely to occur due to their low frequency and possible uneven distribution among populations. It is straightforward to control population stratification in our approach as we can consider population as an environmental factor and apply it to $I_2^*$. An alternative is to treat population with PCA and include the discretized eigenvalues in our analysis.

A major advantage of SPA is that it is highly extensible. The building blocks of SPA are the partitions formed by individual rare variant and it is easy to incorporate complex interactions. As demonstrated in the article, we are able to take into account interactions with environmental factors. Similar approaches can be applied when considering interactions with common variants or other covariates. It can also be generalized to other research areas to benefit the practitioners and scientists in various fields. We believe that the proposed framework of SPA will offer substantial opportunities in detecting potential complicated interactions. Once interaction effects indeed exist, our approach is capable of identifying these interactions and thus adding to the detection power.

This paper presents a simple novel (and easily implemented) tool SPA as an alternative to existing statistical methods for rare variants association studies, with a unique additional feature that SPA can easily incorporate various forms of interaction effects. This addition may add considerable power to disease-related detection in the future. From our studies, if the underlying model is a simple linear additive model with only marginal effects, the powers of SPA are comparable to several existing methods. However, if the model is more complex with interaction effects, the proposed approach provides a more powerful alternative in rare variants association analysis so that there is a better chance to find disease-associated factors. With the development of next-generation sequencing techniques, more and more data with a large amount of rare variants will be generated. It is highly unlikely that the disease phenotype is associated with genetic factors through a simple linear main effect model, so the proposed approach is going to be a powerful and rewarding tool to explore the complicated interaction effects revealed by larger datasets. It is worth noting that any interaction pattern, whether it is linear or nonlinear, can be detected by SPA, since it is model-free and is not subject to any distribution assumptions. Therefore, it is very robust and effective regardless of how the genotype affects the phenotype. The R code of the proposed test scores is available to download at http://www.columbia.edu/~rf2283/Software.html

## Acknowledgement

We thank Prof. Tian Zheng and Dr. Chien-Hsun Huang for their valuable discussions in the preparation of this paper. We also appreciate the reviewers' invaluable comments, which are both insightful and constructive.

# Tables

## Table 1. Type I error estimates of $I_1$, $I_2$ and $p^*$

| | Dichotomous Trait | | | | | |
|---|---|---|---|---|---|---|
| | α=0.05 | | | α=0.01 | | |
| Sample Size | $I_1$ | $I_2$ | $p^*$ | $I_1$ | $I_2$ | $p^*$ |
| 600 | 0.052 | 0.055 | 0.054 | 0.009 | 0.012 | 0.009 |
| 1000 | 0.053 | 0.055 | 0.053 | 0.010 | 0.010 | 0.010 |
| 1500 | 0.048 | 0.049 | 0.049 | 0.007 | 0.007 | 0.007 |
| 2000 | 0.053 | 0.057 | 0.053 | 0.010 | 0.013 | 0.010 |
| | Continuous Trait | | | | | |
| | α=0.05 | | | α=0.01 | | |
| Sample Size | $I_1$ | $I_2$ | $p^*$ | $I_1$ | $I_2$ | $p^*$ |
| 600 | 0.055 | 0.055 | 0.058 | 0.013 | 0.011 | 0.013 |
| 1000 | 0.05 | 0.046 | 0.044 | 0.009 | 0.005 | 0.010 |
| 1500 | 0.061 | 0.048 | 0.061 | 0.015 | 0.011 | 0.010 |
| 2000 | 0.045 | 0.046 | 0.043 | 0.013 | 0.009 | 0.009 |

## Table 2. Type I error estimates of $I_2^*$ in two different null settings

| | Null-1: No G, E or G×E Effects | | | |
|---|---|---|---|---|
| | α=0.05 | | α=0.01 | |
| Sample Size | $I_2^*$-Global | $I_2^*$-Local | $I_2^*$-Global | $I_2^*$-Local |
| 600 | 0.053 | 0.050 | 0.007 | 0.007 |
| 1000 | 0.047 | 0.046 | 0.009 | 0.007 |
| 1500 | 0.045 | 0.048 | 0.008 | 0.009 |
| 2000 | 0.043 | 0.044 | 0.009 | 0.009 |
| | Null-2 : Marginal Environmental Effect only | | | |
| | α=0.05 | | α=0.01 | |
| Sample Size | $I_2^*$-Global [a] | $I_2^*$-Local | $I_2^*$-Global [a] | $I_2^*$-Local |
| 600 | 0.110 | 0.046 | 0.027 | 0.007 |
| 1000 | 0.169 | 0.050 | 0.058 | 0.012 |
| 1500 | 0.239 | 0.047 | 0.087 | 0.011 |
| 2000 | 0.282 | 0.046 | 0.114 | 0.017 |

[a] This is actually the **power** of $I_2^*$-Global in the presence of marginal environmental effect.

**Table 3. Power of two genes in GAW17 dataset**

|  | FLT1 | ARNT |
|---|---|---|
| #Rare NS[b] SNPs | 19 (10 causal) | 9 (5 causal) |
| $I_1$ | 0.865 | 0.345 |
| $I_2$ | 0.505 | 0.05 |
| $p*$ | 0.775 | 0.22 |
| SKAT | 0.82 | 0.235 |
| SKATint | 0.77 | 0.1 |
| RB | 0 | 0.005 |
| VT | 0.88 | 0.025 |
| WS | 0.95 | 0.075 |
| CMC | 0.95 | 0.055 |

[b]NS: nonsynonymous

# Figures

**Figure 1**

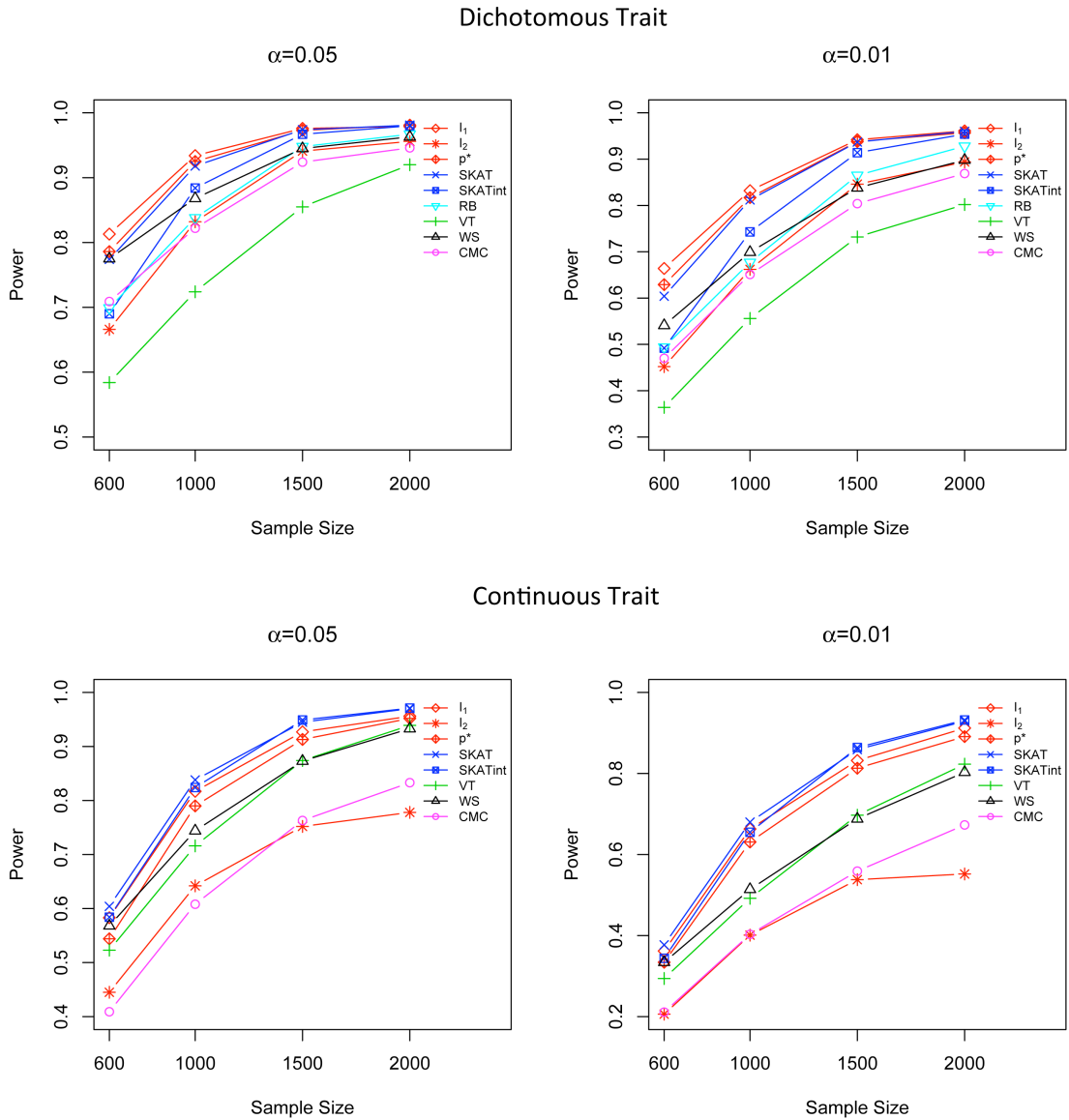

**Figure 1. Power comparison in the marginal effect model when the effect sizes are constant.** Powers were calculated for nominal α levels 0.05 (left) and 0.01(right) and for dichotomous traits (upper) and continuous traits (lower). Powers were evaluated for $I_1$, $I_2$, $p^*$, SKAT, SKATint, VT, RB, WS and CMC. Scenarios with different sample sizes were considered. P-values were estimated using 10,000 permutations and power was evaluated using 1,000 replicates.

**Figure 2**

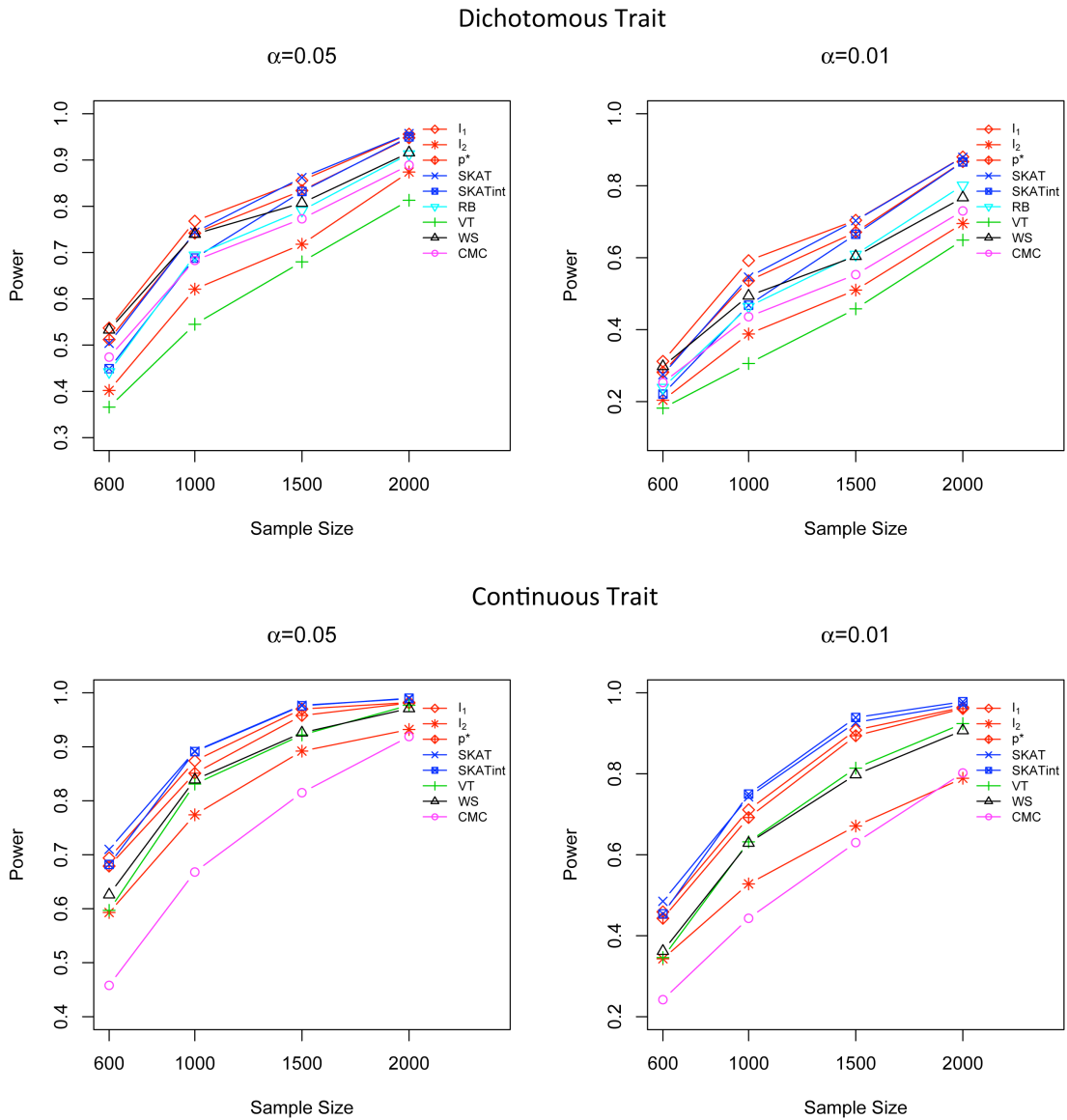

**Figure 2. Power comparison in the marginal effect model when the effect sizes of causal variants are negatively correlated with MAFs.** Powers were calculated for nominal α levels 0.05 (left) and 0.01 (right) and for dichotomous traits (upper) and continuous traits (lower). Powers were evaluated for $I_1$, $I_2$, $p^*$, SKAT, SKATint, VT, RB, WS and CMC. Scenarios with different sample sizes were considered. P-values were estimated using 10,000 permutations and power was evaluated using 1,000 replicates.

**Figure 3**

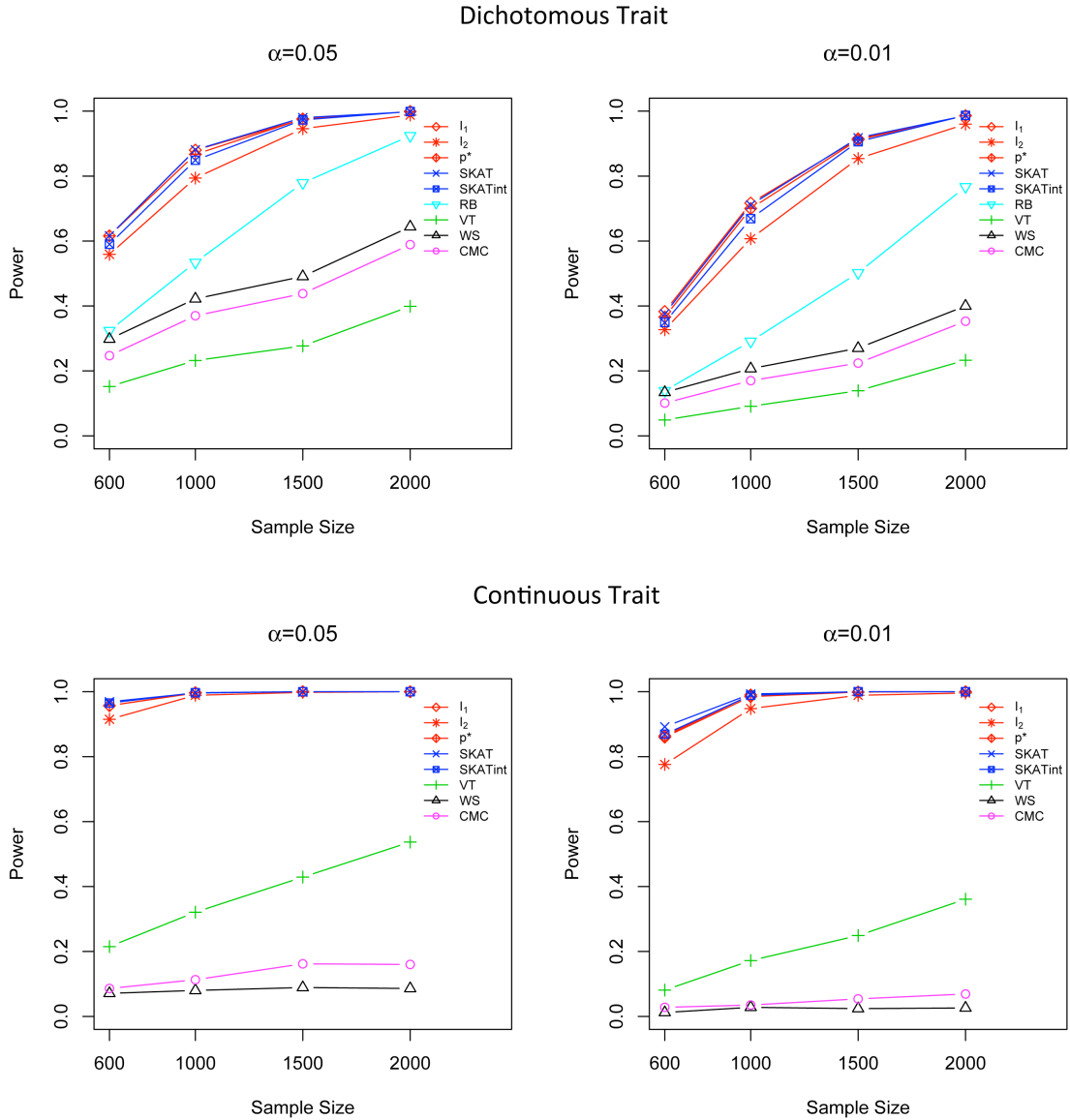

**Figure 3. Power comparison in the marginal effect model with a mixture of protective and risk rare variants.** Powers were calculated for nominal α levels 0.05 (left) and 0.01(right) and for dichotomous traits (upper) and continuous traits (lower). Powers were evaluated for $I_1$, $I_2$, $p^*$, SKAT, SKATint, VT, RB, WS and CMC. Scenarios with different sample sizes were considered. P-values were estimated using 10,000 permutations and power was evaluated using 1,000 replicates.

**Figure 4**

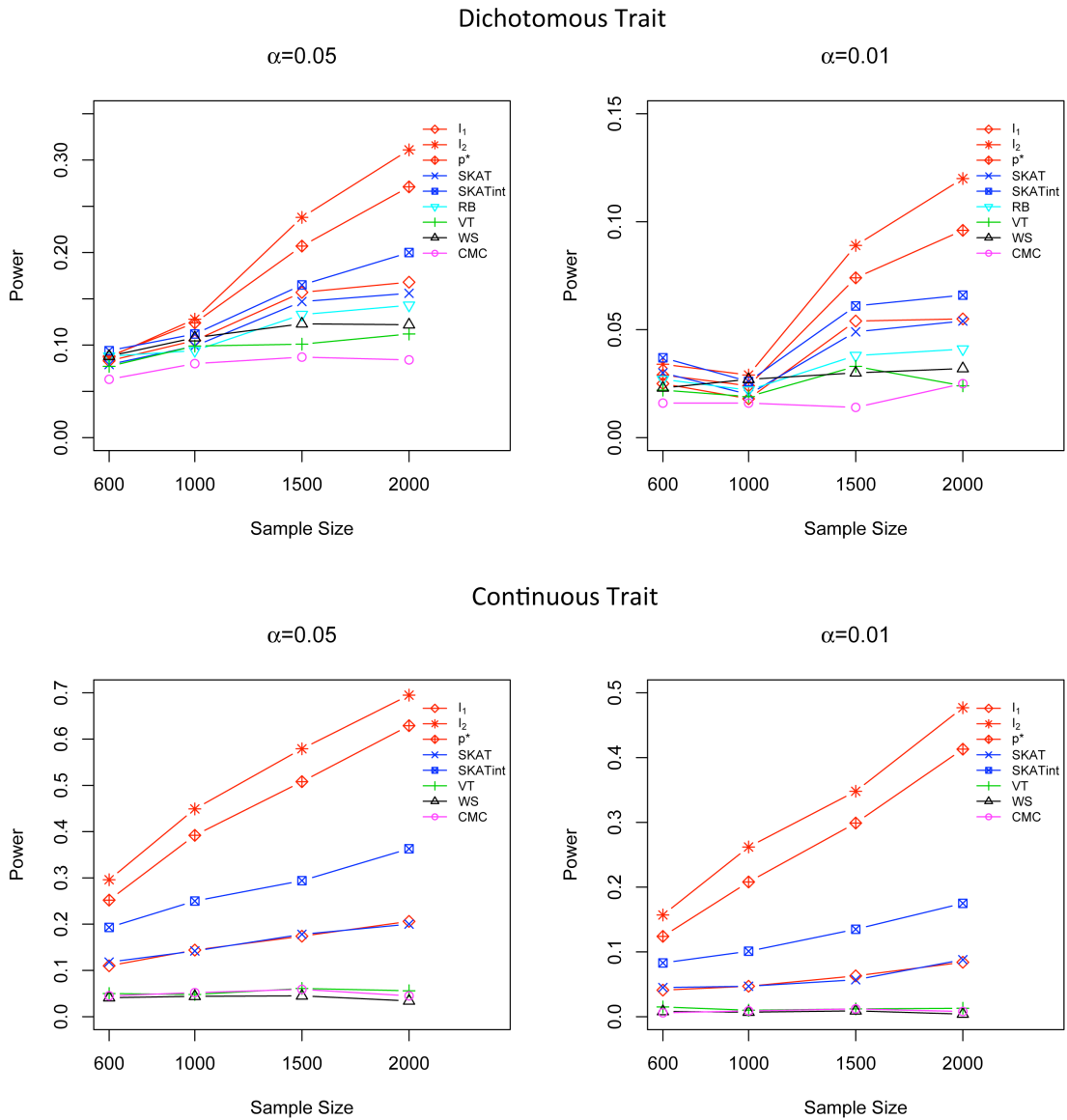

**Figure 4.** Powers are calculated for nominal α levels 0.05 (left) and 0.01(right) and for dichotomous traits (upper) and continuous trait (lower). Power was evaluated for $I_1$, $I_2$, $p^*$, SKAT, SKATint, VT, RB, WS and CMC. Scenarios with different sample sizes were considered. P-values were estimated using 10,000 permutations and power was evaluated using 1,000 replicates.

**Figure 5**

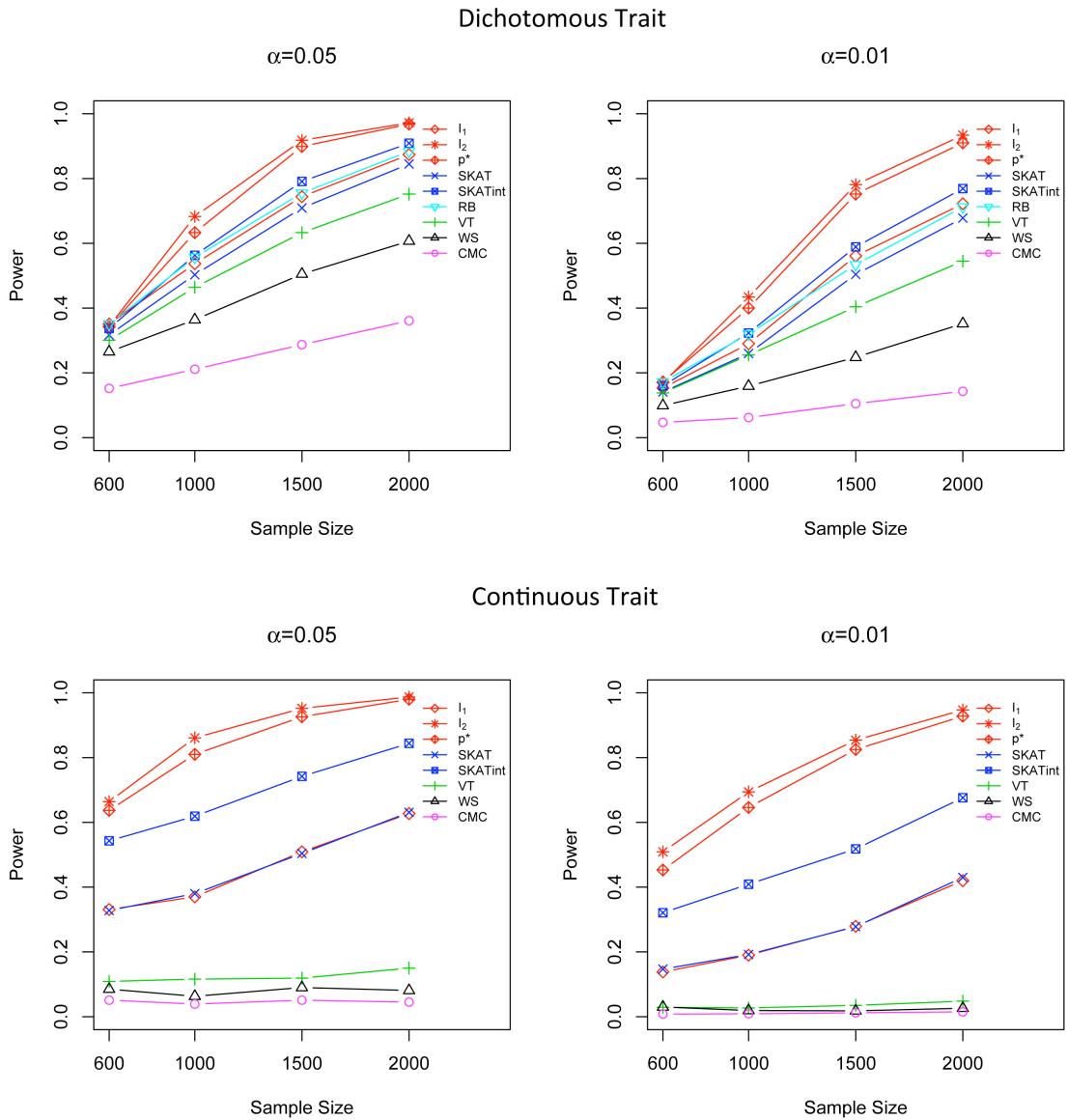

**Figure 5. Power comparison in G×G interaction effect model when 75% of rare variants participate in the interaction effect**. Powers are calculated for nominal α levels 0.05 (left) and 0.01 (right) and for dichotomous traits (upper) and continuous trait (lower). Power was evaluated for $I_1$, $I_2$, $p^*$, SKAT, SKATint, VT, RB, WS and CMC. Scenarios with different sample sizes were considered. P-values were estimated using 10,000 permutations and power was evaluated using 1,000 replicates.

**Figure 6**

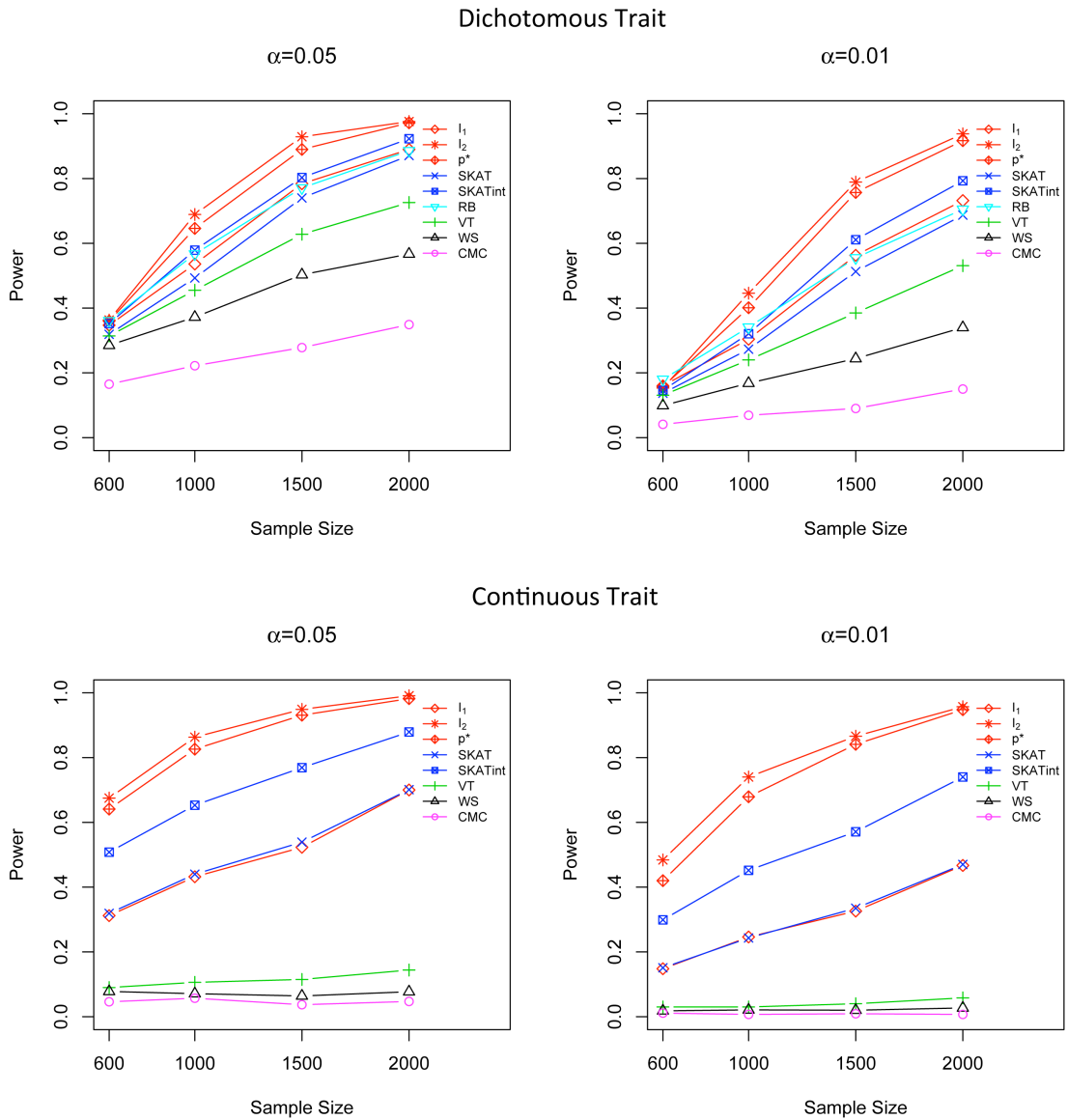

**Figure 6. Power comparison in the scenario with both main effect and G×G interaction effect.** Powers are calculated for nominal α levels 0.05 (left) and 0.01 (right) and for dichotomous traits (upper) and continuous trait (lower). Power was evaluated for $I_1$, $I_2$, $p^*$, SKAT, SKATint, VT, RB, WS and CMC. Scenarios with different sample sizes were considered. P-values were estimated using 10,000 permutations and power was evaluated using 1,000 replicates.

**Figure 7**

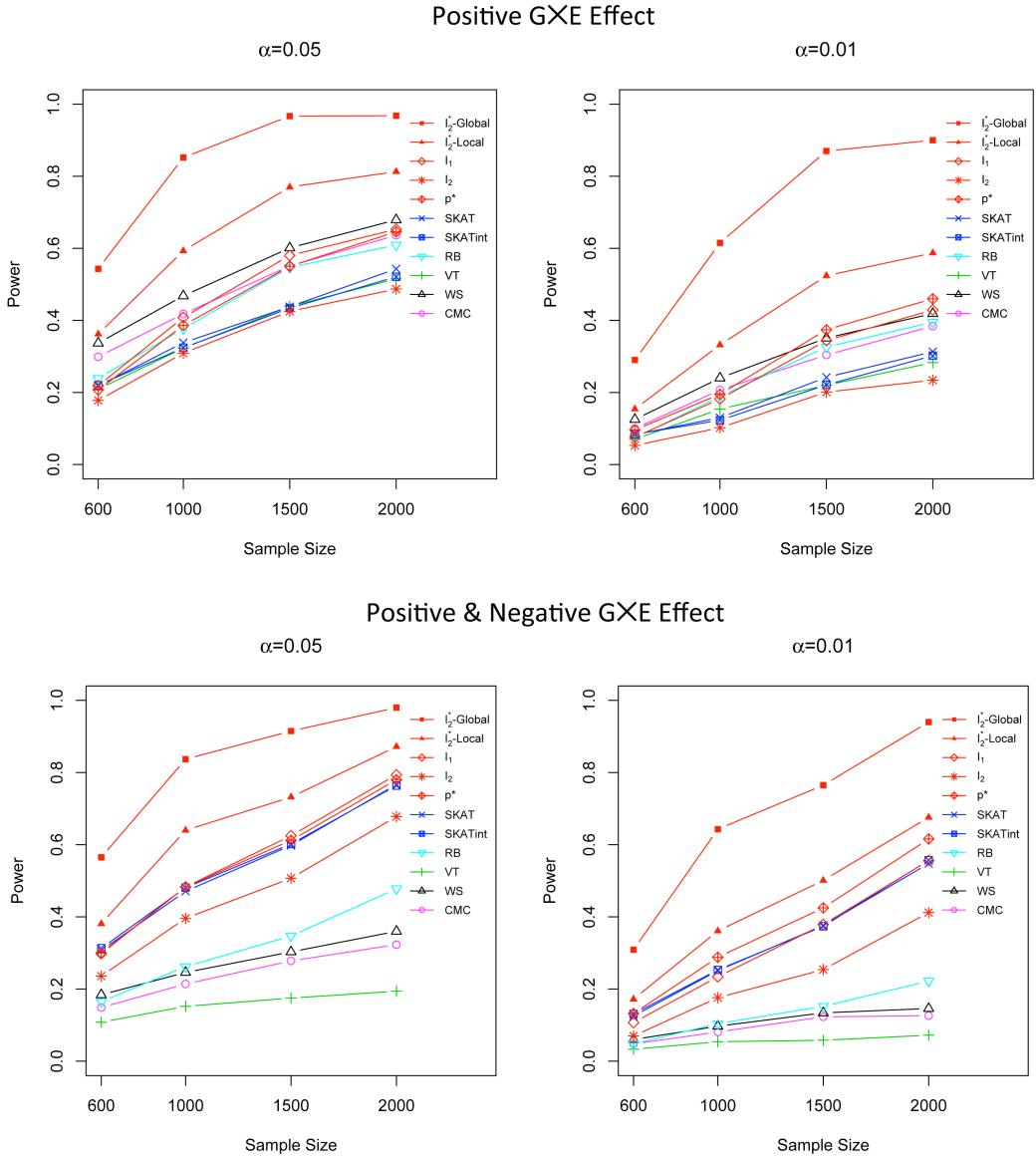

**Figure 7. Power comparison in two G×E interaction models for dichotomous trait.** Powers were calculated for nominal α levels 0.05 (left) and 0.01 (right) when only positive G×E effects exist (upper) and when both positive and negative G×E effects exist (lower). Powers were evaluated for $I_2^*$ (with both global and local permutations), $I_1$, $I_2$, $p^*$, SKAT, SKATint, VT, RB, WS and CMC. Scenarios with different sample sizes were considered. P-values were estimated using 10,000 permutations and power was evaluated using 1,000 replicates.

# Supporting Information

**Supplementary Tables:**

Table S1. Models to generate simulated phenotypes

| | Null Models | |
|---|---|---|
| | Dichotomous | Continuous |
| Null-1 (null setting for $I_1$, $I_2$, $p^*$ and $I_2^*$-Global) | $\log\left(\dfrac{P(A)}{1-P(A)}\right) = \log(1/99)$ | $y = 0.5*U_1 + 0.5*U_2 + \varepsilon$ |
| Null-2 (null setting for $I_2^*$-Local) | $\log\left(\dfrac{P(A)}{1-P(A)}\right) = \log(1/99) + \log(2)*E$ | N/A |
| | Genetic Marginal Effect Models | |
| | Dichotomous | Continuous |
| Scenario 1 (constant marginal effects) | $\log\left(\dfrac{P(A)}{1-P(A)}\right) = \log(1/99) + \log(3)*\sum_{i=1}^{5} X_i$ | $y = 0.5*U_1 + 0.5*U_2 + 0.8*\sum_{i=1}^{5} X_i + \varepsilon$ |
| Scenario 2 (marginal effects negatively correlated with MAF) | $\log\left(\dfrac{P(A)}{1-P(A)}\right) = \log(1/99) + \sum_{i=1}^{5} \beta_i X_i$, where $\beta_i = |\log_{10} MAF|*\ln 5/4$ | $y = 0.5*U_1 + 0.5*U_2 + \sum_{i=1}^{5} \beta_i X_i + \varepsilon$, where $\beta_i = |\log_{10} MAF|*0.4$ |
| Scenario 3 (both positive and negative marginal effect) | $\log\left(\dfrac{P(A)}{1-P(A)}\right) = \log(1/99) + \sum_{i=1}^{5} \beta_i X_i - \sum_{i=6}^{10} \beta_i X_i$, where $\beta_i = |\log_{10} MAF|*\ln 5/4$ | $y = 0.5*U_1 + 0.5*U_2 + \sum_{i=1}^{5} \beta_i X_i - \sum_{i=6}^{10} \beta_i X_i + \varepsilon$, where $\beta_i = |\log_{10} MAF|*0.4$ |
| | G×G Interaction Effect Models | |
| | Dichotomous | Continuous |
| Scenario 4 (50% SNPs participate in G×G interaction effects) | $\log\left(\dfrac{P(A)}{1-P(A)}\right) = \log(1/99) +$ $\sum_{i=1}^{5}\sum_{j=6}^{10} \log 5 * I(X_i > 0, X_j > 0)*(-1)^i$, | $y = 0.5*U_1 + 0.5*U_2 +$ $\sum_{i=1}^{5}\sum_{j=6}^{10} 1.5 * I(X_i > 0, X_j > 0)*(-1)^i + \varepsilon$ |
| Scenario 5 (75% SNPs participate in G×G interaction effects) | $\log\left(\dfrac{P(A)}{1-P(A)}\right) = \log(1/99) +$ $\sum_{i=1}^{6}\sum_{j=7}^{15} \log 5 * I(X_i > 0, X_j > 0)*(-1)^i$, | $y = 0.5*U_1 + 0.5*U_2 +$ $\sum_{i=1}^{6}\sum_{j=7}^{15} 1.5 * I(X_i > 0, X_j > 0)*(-1)^i + \varepsilon$ |

| Scenario 6 (Both marginal and G×G interaction effects) | $\log\left(\dfrac{P(A)}{1-P(A)}\right) = \log\dfrac{1}{99} + \left(\sum_{i=1}^{2}\beta_i X_i - \sum_{i=3}^{5}\beta_i X_i\right)*0.1 + \sum_{i=1}^{6}\sum_{j=7}^{15}\log 5 * I(X_i>0, X_j>0)*(-1)^i$ where $\beta_i = |\log_{10} MAF|*\ln 5/4$ | $y = 0.5*U_1 + 0.5*U_2 + \left(\sum_{i=1}^{2}\beta_i X_i - \sum_{i=3}^{5}\beta_i X_i\right)*0.1 + \sum_{i=1}^{6}\sum_{j=7}^{15} 1.5 * I(X_i>0, X_j>0)*(-1)^i + \varepsilon$ where $\beta_i = |\log_{10} MAF|*0.4$ |
|---|---|---|
| | G×E Interaction Effect Models | |
| | Dichotomous | Continuous |
| Scenario 7 (Positive G×E effect) | $\log\left(\dfrac{P(A)}{1-P(A)}\right) = \log(1/99) + \log(2)*E + \sum_{i=1}^{5}\beta_i * E * X_i$ where $\beta_i = |\log_{10} MAF|*\ln 5/4$ | N/A |
| Scenario 8 (Positive and negative G×E effect) | $\log\left(\dfrac{P(A)}{1-P(A)}\right) = \log(1/99) + \log 2*E + \sum_{i=1}^{5}\beta_i * E * X_i - \sum_{i=6}^{10}\beta_i * E * X_i$ where $\beta_i = |\log_{10} MAF|*\ln 5/4$ | N/A |

We generated 20 independent SNPs with MAF uniformly distributed in 0.0001 and 0.01 except in scenarios 4 and 5 where all the MAFs are set as 0.01. For models involving environmental factors, the environmental factor $E$ is generated from a Bernoulli distribution with success probability 0.5. In models for continuous traits, $U_1 \sim N(0,1)$ and $U_2 \sim Bernoulli(0.5)$ are two covariates independent of genetic factors and $\varepsilon \sim N(0,1)$ is the random noise.

Table S2. Power of different methods for dichotomous traits in scenarios 1~6 ($\alpha$=0.05)

| Scenario | Sample Size | CMC | WS | VT | RB | SKAT | SKATint | $I_1$ | $I_2$ | $p^*$ |
|---|---|---|---|---|---|---|---|---|---|---|
| 1 | 600 | 0.709 | 0.775 | 0.584 | 0.699 | 0.774 | 0.690 | 0.813 | 0.666 | 0.786 |
| 1 | 1000 | 0.822 | 0.868 | 0.724 | 0.838 | 0.918 | 0.884 | 0.934 | 0.832 | 0.925 |
| 1 | 1500 | 0.924 | 0.945 | 0.855 | 0.948 | 0.974 | 0.967 | 0.976 | 0.941 | 0.973 |
| 1 | 2000 | 0.946 | 0.963 | 0.92 | 0.967 | 0.981 | 0.980 | 0.979 | 0.956 | 0.981 |
| 2 | 600 | 0.474 | 0.533 | 0.366 | 0.442 | 0.504 | 0.449 | 0.537 | 0.402 | 0.512 |
| 2 | 1000 | 0.683 | 0.740 | 0.545 | 0.694 | 0.744 | 0.688 | 0.768 | 0.621 | 0.742 |
| 2 | 1500 | 0.773 | 0.807 | 0.680 | 0.791 | 0.862 | 0.832 | 0.856 | 0.718 | 0.834 |
| 2 | 2000 | 0.889 | 0.916 | 0.813 | 0.913 | 0.957 | 0.950 | 0.956 | 0.874 | 0.948 |
| 3 | 600 | 0.247 | 0.298 | 0.152 | 0.324 | 0.616 | 0.590 | 0.616 | 0.559 | 0.616 |
| 3 | 1000 | 0.370 | 0.422 | 0.232 | 0.534 | 0.881 | 0.849 | 0.880 | 0.794 | 0.867 |
| 3 | 1500 | 0.438 | 0.491 | 0.277 | 0.779 | 0.980 | 0.973 | 0.975 | 0.946 | 0.976 |
| 3 | 2000 | 0.589 | 0.644 | 0.399 | 0.924 | 0.998 | 0.999 | 0.999 | 0.988 | 0.999 |
| 4 | 600 | 0.063 | 0.088 | 0.077 | 0.088 | 0.079 | 0.094 | 0.083 | 0.087 | 0.088 |
| 4 | 1000 | 0.080 | 0.108 | 0.099 | 0.094 | 0.099 | 0.112 | 0.105 | 0.128 | 0.124 |
| 4 | 1500 | 0.087 | 0.123 | 0.101 | 0.133 | 0.147 | 0.165 | 0.157 | 0.238 | 0.207 |
| 4 | 2000 | 0.084 | 0.122 | 0.112 | 0.143 | 0.156 | 0.200 | 0.168 | 0.311 | 0.271 |
| 5 | 600 | 0.152 | 0.265 | 0.301 | 0.351 | 0.317 | 0.337 | 0.351 | 0.344 | 0.342 |
| 5 | 1000 | 0.211 | 0.364 | 0.464 | 0.557 | 0.503 | 0.563 | 0.537 | 0.683 | 0.633 |
| 5 | 1500 | 0.287 | 0.505 | 0.633 | 0.755 | 0.709 | 0.791 | 0.744 | 0.918 | 0.899 |
| 5 | 2000 | 0.361 | 0.607 | 0.752 | 0.884 | 0.845 | 0.909 | 0.874 | 0.972 | 0.968 |
| 6 | 600 | 0.165 | 0.285 | 0.314 | 0.361 | 0.319 | 0.352 | 0.347 | 0.363 | 0.360 |
| 6 | 1000 | 0.222 | 0.372 | 0.455 | 0.567 | 0.493 | 0.579 | 0.536 | 0.689 | 0.646 |
| 6 | 1500 | 0.278 | 0.503 | 0.628 | 0.771 | 0.740 | 0.803 | 0.784 | 0.929 | 0.890 |
| 6 | 2000 | 0.349 | 0.567 | 0.726 | 0.886 | 0.871 | 0.923 | 0.890 | 0.976 | 0.972 |

Table S3. Power of different methods for dichotomous traits in scenarios 1~6 ($\alpha=0.01$)

| Scenario | Sample Size | CMC | WS | VT | RB | SKAT | SKATint | $I_1$ | $I_2$ | $p^*$ |
|---|---|---|---|---|---|---|---|---|---|---|
| 1 | 600 | 0.470 | 0.541 | 0.364 | 0.494 | 0.604 | 0.492 | 0.664 | 0.452 | 0.629 |
| 1 | 1000 | 0.651 | 0.699 | 0.556 | 0.677 | 0.812 | 0.743 | 0.832 | 0.662 | 0.817 |
| 1 | 1500 | 0.804 | 0.838 | 0.732 | 0.865 | 0.937 | 0.914 | 0.942 | 0.846 | 0.938 |
| 1 | 2000 | 0.869 | 0.898 | 0.802 | 0.928 | 0.960 | 0.954 | 0.961 | 0.894 | 0.957 |
| 2 | 600 | 0.253 | 0.298 | 0.182 | 0.239 | 0.275 | 0.221 | 0.312 | 0.204 | 0.282 |
| 2 | 1000 | 0.436 | 0.494 | 0.306 | 0.464 | 0.546 | 0.468 | 0.592 | 0.388 | 0.536 |
| 2 | 1500 | 0.553 | 0.604 | 0.458 | 0.609 | 0.703 | 0.665 | 0.704 | 0.510 | 0.671 |
| 2 | 2000 | 0.730 | 0.767 | 0.649 | 0.801 | 0.879 | 0.866 | 0.880 | 0.695 | 0.867 |
| 3 | 600 | 0.101 | 0.134 | 0.049 | 0.139 | 0.376 | 0.349 | 0.383 | 0.327 | 0.366 |
| 3 | 1000 | 0.170 | 0.207 | 0.091 | 0.291 | 0.711 | 0.669 | 0.717 | 0.607 | 0.700 |
| 3 | 1500 | 0.224 | 0.270 | 0.139 | 0.502 | 0.918 | 0.906 | 0.915 | 0.854 | 0.912 |
| 3 | 2000 | 0.353 | 0.400 | 0.233 | 0.767 | 0.986 | 0.987 | 0.986 | 0.960 | 0.986 |
| 4 | 600 | 0.016 | 0.023 | 0.022 | 0.027 | 0.030 | 0.037 | 0.025 | 0.034 | 0.029 |
| 4 | 1000 | 0.016 | 0.027 | 0.019 | 0.022 | 0.020 | 0.026 | 0.018 | 0.029 | 0.024 |
| 4 | 1500 | 0.014 | 0.030 | 0.033 | 0.038 | 0.049 | 0.061 | 0.054 | 0.089 | 0.074 |
| 4 | 2000 | 0.025 | 0.032 | 0.024 | 0.041 | 0.054 | 0.066 | 0.055 | 0.120 | 0.096 |
| 5 | 600 | 0.047 | 0.099 | 0.138 | 0.172 | 0.141 | 0.160 | 0.153 | 0.169 | 0.172 |
| 5 | 1000 | 0.062 | 0.159 | 0.255 | 0.320 | 0.260 | 0.323 | 0.290 | 0.434 | 0.400 |
| 5 | 1500 | 0.105 | 0.248 | 0.404 | 0.534 | 0.504 | 0.589 | 0.561 | 0.781 | 0.752 |
| 5 | 2000 | 0.143 | 0.352 | 0.545 | 0.712 | 0.678 | 0.769 | 0.722 | 0.934 | 0.910 |
| 6 | 600 | 0.041 | 0.099 | 0.131 | 0.180 | 0.137 | 0.146 | 0.160 | 0.153 | 0.157 |
| 6 | 1000 | 0.069 | 0.168 | 0.240 | 0.341 | 0.273 | 0.321 | 0.304 | 0.446 | 0.401 |
| 6 | 1500 | 0.090 | 0.244 | 0.385 | 0.555 | 0.513 | 0.611 | 0.563 | 0.789 | 0.757 |
| 6 | 2000 | 0.150 | 0.340 | 0.531 | 0.705 | 0.687 | 0.793 | 0.732 | 0.938 | 0.917 |

Table S4. Power of different methods for continuous traits in scenarios 1~6 ($\alpha=0.05$)

| Scenario | Sample Size | CMC | WS | VT | SKAT | SKATint | $I_1$ | $I_2$ | $p^*$ |
|---|---|---|---|---|---|---|---|---|---|
| 1 | 600 | 0.409 | 0.568 | 0.523 | 0.604 | 0.584 | 0.583 | 0.445 | 0.544 |
| 1 | 1000 | 0.608 | 0.744 | 0.716 | 0.838 | 0.824 | 0.817 | 0.642 | 0.790 |
| 1 | 1500 | 0.763 | 0.873 | 0.874 | 0.945 | 0.949 | 0.927 | 0.752 | 0.913 |
| 1 | 2000 | 0.833 | 0.933 | 0.939 | 0.970 | 0.971 | 0.956 | 0.778 | 0.952 |
| 2 | 600 | 0.458 | 0.626 | 0.597 | 0.710 | 0.682 | 0.694 | 0.593 | 0.679 |
| 2 | 1000 | 0.668 | 0.839 | 0.831 | 0.892 | 0.891 | 0.874 | 0.774 | 0.851 |
| 2 | 1500 | 0.815 | 0.926 | 0.922 | 0.977 | 0.976 | 0.970 | 0.892 | 0.958 |
| 2 | 2000 | 0.919 | 0.971 | 0.977 | 0.989 | 0.990 | 0.982 | 0.932 | 0.981 |
| 3 | 600 | 0.086 | 0.071 | 0.215 | 0.969 | 0.965 | 0.957 | 0.915 | 0.957 |
| 3 | 1000 | 0.113 | 0.080 | 0.321 | 0.996 | 0.997 | 0.996 | 0.989 | 0.997 |
| 3 | 1500 | 0.162 | 0.089 | 0.429 | 1.000 | 1.000 | 1.000 | 0.998 | 1.000 |
| 3 | 2000 | 0.160 | 0.086 | 0.537 | 1.000 | 1.000 | 1.000 | 1.000 | 1.000 |
| 4 | 600 | 0.045 | 0.041 | 0.050 | 0.118 | 0.193 | 0.110 | 0.296 | 0.252 |
| 4 | 1000 | 0.052 | 0.044 | 0.048 | 0.142 | 0.250 | 0.144 | 0.449 | 0.392 |
| 4 | 1500 | 0.059 | 0.045 | 0.061 | 0.178 | 0.294 | 0.174 | 0.579 | 0.508 |
| 4 | 2000 | 0.045 | 0.034 | 0.056 | 0.200 | 0.363 | 0.206 | 0.695 | 0.629 |
| 5 | 600 | 0.051 | 0.085 | 0.109 | 0.328 | 0.543 | 0.331 | 0.664 | 0.637 |
| 5 | 1000 | 0.039 | 0.063 | 0.116 | 0.380 | 0.619 | 0.370 | 0.861 | 0.810 |

| Scenario | Sample Size | | | | | | | |
|---|---|---|---|---|---|---|---|---|
| 5 | 1500 | 0.051 | 0.090 | 0.119 | 0.504 | 0.742 | 0.509 | 0.952 | 0.926 |
| 5 | 2000 | 0.045 | 0.081 | 0.150 | 0.631 | 0.844 | 0.627 | 0.987 | 0.980 |
| 6 | 600 | 0.046 | 0.078 | 0.090 | 0.319 | 0.508 | 0.312 | 0.675 | 0.641 |
| 6 | 1000 | 0.057 | 0.071 | 0.106 | 0.440 | 0.653 | 0.432 | 0.863 | 0.826 |
| 6 | 1500 | 0.037 | 0.064 | 0.115 | 0.539 | 0.769 | 0.523 | 0.949 | 0.931 |
| 6 | 2000 | 0.047 | 0.077 | 0.144 | 0.701 | 0.879 | 0.700 | 0.991 | 0.982 |

Table S5. Power of different methods for continuous traits in scenarios 1~6 ($\alpha=0.01$)

| Scenario | Sample Size | CMC | WS | VT | SKAT | SKATint | $I_1$ | $I_2$ | $p^*$ |
|---|---|---|---|---|---|---|---|---|---|
| 1 | 600 | 0.211 | 0.334 | 0.294 | 0.377 | 0.344 | 0.362 | 0.206 | 0.334 |
| 1 | 1000 | 0.403 | 0.514 | 0.492 | 0.680 | 0.655 | 0.663 | 0.401 | 0.631 |
| 1 | 1500 | 0.559 | 0.688 | 0.697 | 0.859 | 0.864 | 0.832 | 0.538 | 0.813 |
| 1 | 2000 | 0.673 | 0.803 | 0.823 | 0.929 | 0.932 | 0.912 | 0.552 | 0.891 |
| 2 | 600 | 0.242 | 0.362 | 0.346 | 0.485 | 0.454 | 0.459 | 0.343 | 0.443 |
| 2 | 1000 | 0.443 | 0.629 | 0.632 | 0.743 | 0.750 | 0.711 | 0.528 | 0.692 |
| 2 | 1500 | 0.630 | 0.798 | 0.814 | 0.927 | 0.939 | 0.908 | 0.671 | 0.894 |
| 2 | 2000 | 0.802 | 0.907 | 0.924 | 0.971 | 0.978 | 0.964 | 0.789 | 0.961 |
| 3 | 600 | 0.028 | 0.012 | 0.081 | 0.892 | 0.869 | 0.864 | 0.776 | 0.860 |
| 3 | 1000 | 0.035 | 0.028 | 0.172 | 0.993 | 0.989 | 0.990 | 0.948 | 0.985 |
| 3 | 1500 | 0.054 | 0.024 | 0.249 | 1.000 | 1.000 | 0.999 | 0.989 | 0.999 |
| 3 | 2000 | 0.069 | 0.026 | 0.361 | 1.000 | 1.000 | 1.000 | 0.996 | 1.000 |
| 4 | 600 | 0.006 | 0.008 | 0.015 | 0.045 | 0.083 | 0.041 | 0.157 | 0.124 |
| 4 | 1000 | 0.009 | 0.007 | 0.010 | 0.047 | 0.101 | 0.047 | 0.262 | 0.208 |
| 4 | 1500 | 0.012 | 0.009 | 0.012 | 0.057 | 0.135 | 0.063 | 0.348 | 0.299 |
| 4 | 2000 | 0.008 | 0.004 | 0.013 | 0.088 | 0.175 | 0.084 | 0.477 | 0.413 |
| 5 | 600 | 0.008 | 0.030 | 0.029 | 0.147 | 0.321 | 0.138 | 0.509 | 0.453 |
| 5 | 1000 | 0.009 | 0.019 | 0.027 | 0.192 | 0.409 | 0.190 | 0.694 | 0.646 |
| 5 | 1500 | 0.012 | 0.018 | 0.035 | 0.278 | 0.518 | 0.279 | 0.854 | 0.825 |
| 5 | 2000 | 0.015 | 0.026 | 0.048 | 0.430 | 0.676 | 0.420 | 0.947 | 0.928 |
| 6 | 600 | 0.011 | 0.018 | 0.030 | 0.151 | 0.299 | 0.148 | 0.484 | 0.420 |
| 6 | 1000 | 0.007 | 0.021 | 0.030 | 0.243 | 0.452 | 0.246 | 0.740 | 0.679 |
| 6 | 1500 | 0.009 | 0.020 | 0.040 | 0.335 | 0.571 | 0.326 | 0.866 | 0.841 |
| 6 | 2000 | 0.007 | 0.027 | 0.058 | 0.470 | 0.740 | 0.467 | 0.957 | 0.948 |

Table S6. Power of different methods for dichotomous traits in G×E interaction effect models ($\alpha=0.05$)

| Scenario | Sample Size | CMC | WS | VT | RB | SKAT | SKATint | $I_1$ | $I_2$ | $p^*$ | $I_2^*$-Global | $I_2^*$-Local |
|---|---|---|---|---|---|---|---|---|---|---|---|---|
| 7 | 600 | 0.299 | 0.337 | 0.209 | 0.239 | 0.218 | 0.220 | 0.219 | 0.178 | 0.207 | 0.543 | 0.362 |
| 7 | 1000 | 0.418 | 0.468 | 0.323 | 0.377 | 0.338 | 0.324 | 0.408 | 0.309 | 0.386 | 0.852 | 0.593 |
| 7 | 1500 | 0.551 | 0.601 | 0.439 | 0.547 | 0.438 | 0.434 | 0.580 | 0.425 | 0.550 | 0.967 | 0.770 |
| 7 | 2000 | 0.636 | 0.679 | 0.515 | 0.609 | 0.543 | 0.521 | 0.653 | 0.487 | 0.646 | 0.968 | 0.813 |
| 8 | 600 | 0.149 | 0.184 | 0.108 | 0.165 | 0.307 | 0.313 | 0.298 | 0.236 | 0.300 | 0.565 | 0.381 |
| 8 | 1000 | 0.214 | 0.246 | 0.152 | 0.262 | 0.471 | 0.481 | 0.483 | 0.396 | 0.483 | 0.837 | 0.640 |
| 8 | 1500 | 0.278 | 0.303 | 0.175 | 0.347 | 0.598 | 0.602 | 0.625 | 0.507 | 0.612 | 0.915 | 0.732 |
| 8 | 2000 | 0.323 | 0.360 | 0.194 | 0.478 | 0.767 | 0.764 | 0.794 | 0.678 | 0.780 | 0.980 | 0.872 |

Table S7. Power of different methods for dichotomous traits in G×E interaction effect models (α=0.01)

| Scenario | Sample Size | CMC | WS | VT | RB | SKAT | SKATint | $I_1$ | $I_2$ | $p^*$ | $I_2^*$-Global | $I_2^*$-Local |
|---|---|---|---|---|---|---|---|---|---|---|---|---|
| 7 | 600 | 0.100 | 0.125 | 0.07 | 0.074 | 0.083 | 0.084 | 0.076 | 0.053 | 0.096 | 0.29 | 0.154 |
| 7 | 1000 | 0.207 | 0.240 | 0.154 | 0.189 | 0.130 | 0.123 | 0.182 | 0.102 | 0.195 | 0.615 | 0.332 |
| 7 | 1500 | 0.304 | 0.351 | 0.220 | 0.326 | 0.242 | 0.221 | 0.344 | 0.201 | 0.374 | 0.870 | 0.524 |
| 7 | 2000 | 0.384 | 0.419 | 0.283 | 0.395 | 0.313 | 0.302 | 0.429 | 0.234 | 0.460 | 0.900 | 0.587 |
| 8 | 600 | 0.048 | 0.060 | 0.033 | 0.048 | 0.125 | 0.131 | 0.107 | 0.070 | 0.132 | 0.309 | 0.172 |
| 8 | 1000 | 0.081 | 0.097 | 0.054 | 0.104 | 0.251 | 0.253 | 0.234 | 0.176 | 0.288 | 0.643 | 0.361 |
| 8 | 1500 | 0.123 | 0.134 | 0.058 | 0.152 | 0.376 | 0.374 | 0.379 | 0.254 | 0.425 | 0.765 | 0.501 |
| 8 | 2000 | 0.126 | 0.146 | 0.072 | 0.222 | 0.548 | 0.557 | 0.556 | 0.412 | 0.616 | 0.940 | 0.676 |

**Supplementary Figures**

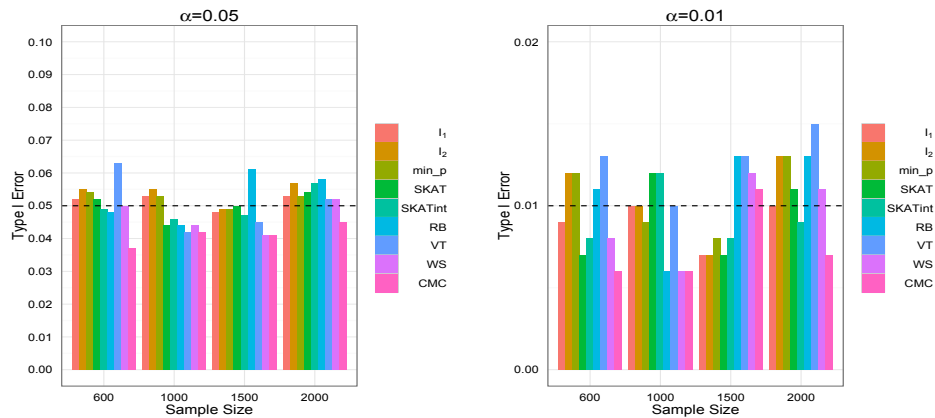

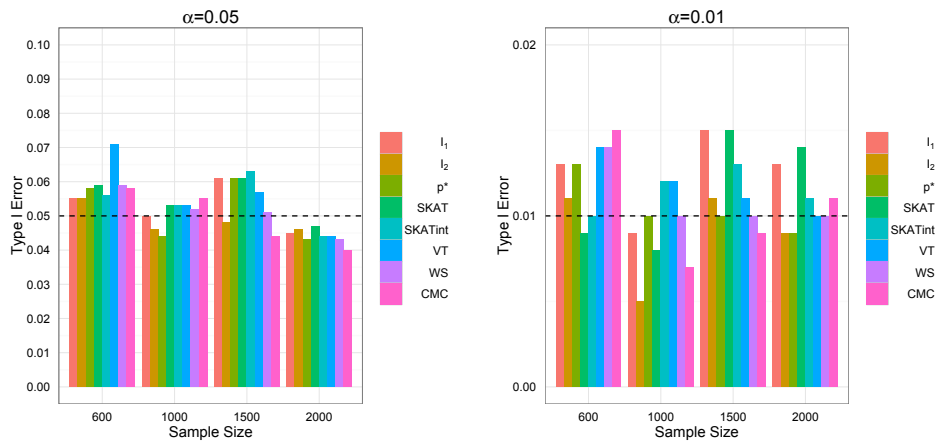

**Figure S1. Type I error for different methods in various sample sizes with nominal α levels 0.05 (left) and 0.01(right).** Results for four sample sizes: 600, 1000, 1500 and 2000, with equal numbers of cases and controls.

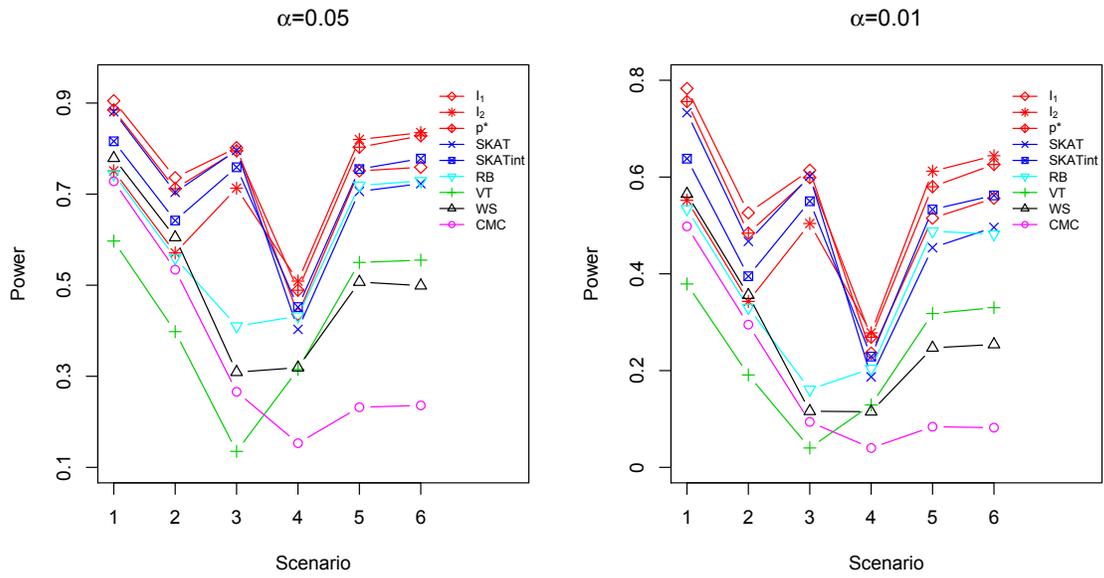

**Figure S2. Power comparison in scenarios 1~6 for dichotomous traits with 500 cases and 500 controls when the SNP number is 30.** Powers were calculated for nominal α levels 0.05 (left) and 0.01(right). Power was evaluated for $I_1$, $I_2$, $p^*$, SKAT, SKATint, VT, RB, WS and CMC. P-values were estimated using 10,000 permutations and power was evaluated using 1,000 replicates.